\documentclass{ws-ijmpcs}

\begin{document}

\markboth{C. Furtado, T. Mariz, J. R. Nascimento, A. Yu. Petrov, A. F. Santos}
{The G\"{o}del metric in the Chern-Simons modified gravity}

%
\catchline{}{}{}{}{}
%

\title{THE G\"{O}DEL METRIC IN THE CHERN-SIMONS MODIFIED GRAVITY}

\author{C. FURTADO}

\address{
Departamento de F\'{\i}sica, Universidade Federal da Para\'{\i}ba\\
Caixa Postal 5008, 58051-970, Jo\~ao Pessoa, Para\'{\i}ba, Brazil\\
furtado@fisica.ufpb.br}

\author{T. MARIZ}
\address{Instituto de F\'{\i}sica, Universidade Federal de Alagoas\\ 
57072-970, Macei\'{o}, Alagoas, Brazil\\
tmariz@if.ufal.br}

\author{J. R. NASCIMENTO}

\address{
Departamento de F\'{\i}sica, Universidade Federal da Para\'{\i}ba\\
Caixa Postal 5008, 58051-970, Jo\~ao Pessoa, Para\'{\i}ba, Brazil\\
jroberto@fisica.ufpb.br}

\author{A. YU. PETROV}

\address{
Departamento de F\'{\i}sica, Universidade Federal da Para\'{\i}ba\\
Caixa Postal 5008, 58051-970, Jo\~ao Pessoa, Para\'{\i}ba, Brazil\\
petrov@fisica.ufpb.br}

\author{A. F. SANTOS}

\address{
Instituto de F\'{\i}sica, Universidade Federal de Mato Grosso,\\
78060-900, Cuiab\'{a}, Mato Grosso, Brazil\\
alesandroferreira@fisica.ufmt.br}

\maketitle

\begin{history}
\received{Day Month Year}
\revised{Day Month Year}
\comby{Managing Editor}
\end{history}

\begin{abstract}
We discuss the consistency  of the G\"{o}del metric within the Chern-Simons modified gravity, both for external and dynamical Chern-Simons coefficients.
\end{abstract}


The four-dimensional Chern-Simons (CS) modified gravity has been recently intensively studied within different contexts. The main reason for this consists in the fact that this theory, being a straightforward generalization of the three-dimensional CS modified gravity\cite{DJT}, and formulated initially in\cite{JaPi} can be naturally treated as a simple example of the CPT breaking gravity theory. Further, a lot of interesting results has been obtained for the CS modified gravity, such as birefringence of the gravitational waves within this model\cite{JaPi}, modification of orbits around the Earth due to the gravitational CS term\cite{Smith}, post-Newtonian expansion\cite{Alexander}, cosmological impacts\cite{Lue,Alexander01}. In\cite{ours} the dynamical generation of the gravitational CS term via perturbative corrections was carried out.  

One of the principal lines of study of the CS modified gravity is a study of consistency of the known spacetime solutions within this theory. Already in\cite{JaPi} this compatibility has been shown for the Schwarzschild metric. Further, in\cite{Grumiller} it was proved that the solutions of the usual Einstein equations possessing certain symmetries, in particular, spherically symmetric and axisymmetric ones, solve the equations of motion of the CS modified gravity as well. 

Another important solution is the G\"{o}del metric\cite{Godel}, which is the first known metric allowing for the breaking of causality through possibility of the closed timelike curves (CTCs). A lot of issues related to this metric has been considered earlier\cite{Reb,ourGodel}, including, in particular, constructing of its generalizations and study of the possibilities of arising the CTCs within these generalizations. Therefore, the natural question is -- whether the  G\"{o}del metric\cite{Godel} is consistent within the CS modified gravity? In other words, whether the CS modified gravity allows for existence of the CTCs?
In this paper we are interested in verifying whether the G\"{o}del solution holds in the CS modified gravity, both for external and dynamical CS terms.

We start our study with the introduction of the G\"{o}del metric which is written as\cite{Godel}
\begin{eqnarray}
ds^2=a^2\Bigl[dt^2-dx^2+\frac{1}{2}e^{2x}dy^2-dz^2+2 e^x dt\,dy\Bigl],\label{godel}
\end{eqnarray}
where $a$ is a positive number. The non-zero Christoffel symbols corresponding to this metric look like 
\begin{eqnarray}
\Gamma^0_{01}=1,\,\,\,\,\,\,\, \Gamma^0_{12}=\frac{1}{2}e^x,\,\,\,\,\,\,\,\Gamma^1_{02}=\frac{1}{2}e^x, \,\,\,\,\,\,\,\Gamma^1_{22}=\frac{1}{2}e^{2x},\,\,\,\,\,\,\,\Gamma^2_{01}=-e^{-x}.
\end{eqnarray}
The non-zero components of the Riemann tensor are
\begin{eqnarray}
R_{0101}=-\frac{1}{2}a^2, \,\,\,R_{0112}=\frac{1}{2}a^2e^x, \,\,\, R_{0202}=-\frac{1}{4}a^2e^{2x}, \,\,\, R_{1212}=-\frac{3}{4}a^2e^{2x}.
\end{eqnarray}
The corresponding non-zero components of the Ricci tensor look like
\begin{eqnarray}
 R_{00}=1, \,\,\,\,\,\,\,\, R_{02}=R_{20}=e^x, \,\,\,\,\,\,\,\, R_{22}=e^{2x}.
\end{eqnarray}
Finally, the Ricci scalar is
\begin{eqnarray}
R=\frac{1}{a^2}.
\end{eqnarray}
It is known\cite{Godel} that the G\"{o}del metric solves the Einstein equations
\begin{eqnarray}
\label{ein}
R_{\mu\nu}-\frac{1}{2}g_{\mu\nu}R=8\pi G\rho u_\mu u_\nu+\Lambda g_{\mu\nu},
\end{eqnarray} 
where $u$ is a unit time-like vector of the form $u^{\mu}=(\frac{1}{a},0,0,0)$, and its covariant components are $u_\mu=(a,0,ae^x,0)$, if
\begin{eqnarray}
\Lambda=-\frac{1}{2a^2},\,\,\,\,\,\,\,\,\,\,\,\,8\pi G\rho=\frac{1}{a^2}.
\end{eqnarray}

Now, let us modify the gravity action by adding the gravitational CS term\cite{JaPi}.
The resulting CS modified gravity action looks like
\begin{eqnarray}
\label{smod}
S=\frac{1}{16\pi G}\int d^4x\Bigl[\sqrt{-g}(R-\Lambda)+\frac{l}{4}\Theta \,{^*}RR\Bigl] + S_{mat},
\end{eqnarray}
where $R$ is the scalar curvature, $S_{mat}$ is the matter action and ${^*}RR$ is the Pontryagin term:
\begin{eqnarray}
{^*}RR\equiv {^*}{R^\alpha}\,_\beta\,^{\gamma\delta}R^\beta\,_{\alpha\gamma\delta}=\frac{1}{2}\epsilon^{\gamma\delta\mu\nu}R^\alpha\,_{\beta\mu\nu}R^\beta\,_{\alpha\gamma\delta},
\end{eqnarray}
where $R^\beta\,_{\alpha\gamma\delta}$ is the Riemann tensor.
The function $\Theta$ is a scalar field called the CS coefficient. It can be treated as an external one or dynamical one. Within this paper, we consider both these possibilities.

After integration by parts, introducing $v_{\mu}=\partial_{\mu}\Theta$, we get
\begin{eqnarray}
\frac{l}{4}\int d^4x \Theta{^*}RR=-l\int d^4x v_{\mu}\epsilon^{\mu\alpha\beta\gamma}(\frac{1}{2}\Gamma^{\sigma}_{\alpha\tau}\partial_{\beta}\Gamma^{\tau}_{\gamma\sigma}+\frac{1}{3}
\Gamma^{\sigma}_{\alpha\tau}\Gamma^{\tau}_{\beta\eta}\Gamma^{\eta}_{\gamma\sigma}),
\end{eqnarray}
that is, the usual gravitational CS term.
Varying the action (\ref{smod}) with respect to the metric, we obtain the modified Einstein equations
\begin{eqnarray}
 R^{\mu\nu}-\frac{1}{2}g^{\mu\nu}R+lC^{\mu\nu}=T^{\mu\nu}, \label{einst}
\end{eqnarray}
where $C^{\mu\nu}$ is the Cotton tensor arising due to the varying of the additive CS term, its explicit form is\cite{DJPi}
\begin{eqnarray}
C^{\mu\nu}=-\frac{1}{2\sqrt{-g}}\Bigl[v_\sigma \epsilon^{\sigma\mu\alpha\beta}D_\alpha R^\nu_\beta+\frac{1}{2}v_{\sigma\tau}\epsilon^{\sigma\nu\alpha\beta}R^{\tau\mu}\,_{\alpha\beta}\Bigl] \,+\, (\mu\longleftrightarrow\nu), 
\end{eqnarray}
with $v_{\sigma\tau}\equiv D_\sigma v_\tau$.  The covariant divergence of this tensor is
\begin{eqnarray}
D_\mu C^{\mu\nu}=\frac{1}{8\sqrt{-g}}v^{\nu}{^*}RR.
\end{eqnarray}
First, let us suggest that $\Theta$ is the external field. Thus, it does not contribute to the energy-momentum tensor of the matter, which hence has the same form as in the usual case\cite{Godel}:
$T^{\mu\nu}=8\pi G\rho u^\mu u^\nu+\Lambda g^{\mu\nu}$, that is, just the r.h.s. of the Eq. (\ref{ein}) 
Using the Bianchi identity, $D_\mu G^{\mu\nu}=0$, and suggesting that the matter terms are diffeomorphism invariant, i.e. $D_\mu T^{\mu\nu}=0$, one finds that the solution of the equation (\ref{einst}) requires a consistency condition ${^*}RR=0$,
i.e.  the diffeomorphism symmetry breaking is suppressed on-shell (for more details see\cite{JaPi}).

As the G\"{o}del metric solves the usual Einstein equations, it can solve the modified ones if and only if the Cotton tensor vanishes for such a metric. A straightforward analysis shows that this situation takes place only if
\begin{eqnarray}
\label{thf}
\Theta=F(x,y).
\end{eqnarray}
Therefore, for this specific choice of the CS coefficient, we find that the G\"{o}del metric is compatible with the CS modified gravity. Therefore, the possibility for existence of the CTCs  persists in the CS modified gravity. 

Now, let us suggest that the $\Theta$ is dynamical. In this case, the action of the CS modified gravity is 
\begin{eqnarray}
\label{smod1}
S=\frac{1}{16\pi G}\int d^4x\Bigl[\sqrt{-g}(R-\Lambda)+\frac{l}{4}\Theta \,{^*}RR-\frac{1}{2}\partial^\mu\Theta\partial_\mu\Theta\Bigl] + S_{mat}.
\end{eqnarray}
The corresponding equations of motion look like
\begin{eqnarray}
&& G_{\mu\nu}+lC_{\mu\nu}= T_{\mu\nu}, \quad\,
g^{\mu\nu}\nabla_{\mu}\nabla_{\nu}\Theta=-\frac{l}{64\pi} {^*}RR.
 \label{einst1}
\end{eqnarray}

The energy-momentum tensor $T_{\mu\nu}$ is now a sum of two terms: 
\begin{eqnarray}
T_{\mu\nu}=T_{\mu\nu}^m+T_{\mu\nu}^{\Theta}, 
\end{eqnarray}
where $T_{\mu\nu}^m$ is the energy-momentum tensor of an usual matter with a cosmological term given by  the r.h.s. of the Eq. (\ref{ein}).  Further, the energy-momentum tensor for the $\Theta$ field has the usual form corresponding to the scalar field in a curved space:
\begin{eqnarray}
\label{tm}
T_{\mu\nu}^{\Theta}=(\partial_{\mu}\Theta)(\partial_{\nu}\Theta)-\frac{1}{2}g_{\mu\nu}(\partial^{\lambda}\Theta)
(\partial_{\lambda}\Theta).
\end{eqnarray}

Now, let us suggest that $\Theta$ has the form (\ref{thf}), that is, the Cotton tensor vanishes. In this case,
knowing the Ricci tensor and the scalar curvature for the G\"{o}del metric\cite{Godel},
one can write down the nontrivial components of the Einstein equations, 00, 11, 22 and 33 respectively:
\begin{eqnarray}
\frac{1}{2}&=&8\pi G\rho a^2+\Lambda a^2+\left[\frac{1}{2}(\partial_1\Theta)^2+e^{-2x}(\partial_2\Theta)^2\right],\nonumber\\
\frac{1}{2}&=&-\Lambda a^2+\left[\frac{1}{2}(\partial_1\Theta)^2-e^{-2x}(\partial_2\Theta)^2\right],\nonumber\\
\frac{3}{2}&=&16\pi G\rho a^2+\Lambda a^2+\left[\frac{1}{2}(\partial_1\Theta)^2+e^{-2x}(\partial_2\Theta)^2\right],\nonumber\\
\frac{1}{2}&=&-\Lambda a^2-\left[\frac{1}{2}(\partial_1\Theta)^2+e^{-2x}(\partial_2\Theta)^2\right].\label{33}
\end{eqnarray}
The equations for the components 00 and 02 turn out to coincide identically.  Moreover, the component 12 of modified Einstein equations is
\begin{eqnarray}
(\partial_1\Theta)(\partial_2\Theta)=0.\label{12}
\end{eqnarray}
One can verify that the G\"{o}del metric is compatible only with the trivial potential, at least if we want to preserve the condition of vanishing the Cotton tensor.

A straightforward inspection of these components allows to conclude that solutions of the Einstein equations can be formally written as
\begin{eqnarray}
8\pi G \rho=\frac{1}{a^2};\,\,\,\,\,\,\,\,\,\,\,\,\,\,\,\,\,\, \,\,\,\,\,\,\,\,\,\,\,\,\,\,\,\,\,\,\Lambda=-\frac{1}{2a^2}-\frac{\Sigma}{a^2},\label{17}
\end{eqnarray}
where
\begin{eqnarray}
\Sigma=\frac{1}{2}\left(\partial_1\Theta\right)^2+e^{-2x}\left(\partial_2\Theta\right)^2.
\end{eqnarray}
Therefore, the consistency of the equations of motion requires the cosmological constant not to be constant but, instead, to be a function of spatial coordinates. In principle, spatial variability of the cosmological constant has been discussed\cite{Massa}. However, this scenario seems to be rather exotic, in particular, because it violates homogeneity of space. 

Let us discuss our results. We showed that the situation with the compatibility of the G\"{o}del metric with the CS modified gravity is essentially different in the case of the external CS coefficient and dynamical CS coefficient. Indeed, while in the first case the compatibility is achieved for the special value of $\Theta$, in the second case the cancellation of the Cotton tensor is not sufficient -- one must have the variable cosmological constant which violates the principle of homogeneity of space and conservation of the energy-momentum tensor. 

\section*{Acknowledgments}

Authors are grateful to Profs. M. Reboucas and J. A. S. Lima for important discussions. The work has been supported by Conselho Nacional de Desenvolvimento Cient\'\i fico e Tecnol\'ogico (CNPq) and by CNPq/PRONEX/FAPESQ. A. Yu. P. has been supported by the CNPq project No. 303461-2009/8. A. F. S. has been supported by the CNPq project No. 473571/2010-2.


\end{document}